\documentclass[aps,prd,twocolumn,showpacs,superscriptaddress, floatfix]{revtex4-2}  
\usepackage{graphicx}  
\usepackage{dcolumn}   
\usepackage{bm}        
\usepackage{amssymb}   
\usepackage{booktabs}
\usepackage{hyperref}
\usepackage{xspace}
\usepackage[caption=false]{subfig}

\usepackage[caption=false]{subfig}
\usepackage{xcolor}
\usepackage{subfig}
\usepackage{amsmath}
\usepackage{comment}
\usepackage{subfiles}
\usepackage{placeins}
\usepackage{array}
\usepackage[normalem]{ulem}
\usepackage{xcolor}
\usepackage{float}
\usepackage{todonotes}
\usepackage{dblfloatfix}
\usepackage{multirow}
\usepackage{nicematrix}
\usepackage{xspace}
\newcounter{todocounter}

\newcommand{\mev}{MeV\xspace}

\newcommand{\dt}{\ensuremath{\Delta \theta_{T}}\xspace}
\newcommand{\ptmr}{\ensuremath{\vec{p}_{T\mathrm{,miss,reco}}}\xspace}
\newcommand{\ptn}{\ensuremath{\vec{p}_{T\mathrm{,n}}}\xspace}
\newcommand{\pn}{\ensuremath{\vec{p}_\mathrm{n}}\xspace}
\newcommand{\Tn}{\ensuremath{\mathrm{T}_\mathrm{n}}\xspace}
\newcommand{\Tnr}{\ensuremath{\mathrm{T}_\mathrm{n,reco}}\xspace}
\newcommand{\Tnt}{\ensuremath{\mathrm{T}_\mathrm{n,true}}\xspace}
\newcommand{\Rn}{\ensuremath{\vec{R}_\mathrm{n}}\xspace}
\newcommand{\Rtn}{\ensuremath{\vec{R}_{T\mathrm{,n}}}\xspace}

\newcolumntype{L}[1]{>{\raggedright\hspace{0.5cm}\arraybackslash}p{#1}}
\newcolumntype{C}[1]{>{\centering\hspace{0.5cm}\arraybackslash}p{#1}}
\newcolumntype{R}[1]{>{\raggedleft\hspace{0.5cm}\arraybackslash}p{#1}}
\newcolumntype{P}{>{\hspace{0.3em}}c<{\hspace{0.3em}}}
\newcolumntype{Q}{l<{\hspace{0.3em}}}

\begin{document}

\title{Facilitating neutron and energy reconstruction in neutrino events using the direction of tagged neutrons}
\author{S. Manly, M. Lahare, H. Utaegbulam, D. Last } \affiliation{Department of Physics and Astronomy, University of Rochester, Rochester, New York 14627 USA}
\date{February 6, 2025}

\begin{abstract} 
To date, accelerator neutrino experiments have had only limited success including neutrons in the neutrino event reconstruction event by event.
Produced neutrons often interact in a detector leaving an energy deposition that does not correlate strongly with the kinetic energy of the neutron.
This work explores the inclusion of the direction of a neutron tagged by such an energy deposition in concert with the missing transverse momentum of the reconstructed particles to determine the approximate kinetic energy of the neutron and modify the reconstruction of the energy of the incoming neutrino.
The technique significantly increases the neutron kinetic energy estimation relative to one that assigns the neutron kinetic energy by enforcing transverse momentum balance alone.
When included in the neutrino energy calculation, the neutrino energy resolution is improved and the reconstructed energy is distributed more symmetrically around the true value.
The technique shows promise and might be used to good effect to analyze data taken with experiments able to tag neutron energy deposits with good neutron direction resolution such as
the T2K near detector and the liquid argon detectors in the SBN program and DUNE.  
\end{abstract}

\maketitle


\section{Introduction}

The capability to reconstruct neutrons on an event-by-event basis in accelerator neutrino interactions opens up a number of interesting physics possibilities.
Reconstructed neutrons can be used to improve the neutrino energy reconstruction, flag the occurrence of final-state nuclear interactions or interactions off nucleons correlated in the initial state, analyze transverse kinematic variables in antineutrino interactions, enrich samples of antineutrino interactions on hydrogen nuclei in a hydrocarbon target (possibly leading to an improved measure of the weak axial form factor), and provide another degree of freedom along with new event morphologies to improve and tune neutrino interaction models. 

Neutrino experiments are able to tag thermal neutrons efficiently through nuclear capture.
Some examples of this include the SoLID experiment using $^{6}$Li~\cite{SoLid:2017ema}, Super-Kamiokande~\cite{Super-Kamiokande:2021the} and ANNIE~\cite{ANNIE:2017nng} using Gd, and the Cowan-Reines experiment which first detected reactor neutrinos through the use of neutron capture on Cd~\cite{Reines:1960pr}.
If the event rate is low enough, such tagging can be correlated to a given neutrino interaction. However, a capture tag yields no  information on the neutron initial kinetic energy, \Tn , and the directional information is smeared.
This limits its usefulness as it is necessary to determine both the direction of the neutron and \Tn for inclusion in event-by-event reconstruction.

Historically, neutrino experiments have been largely blind to neutrons with respect to event-by-event reconstruction.  Recently, the MINERvA experiment succeeded in associating neutrons with particular events from neutrino interactions on a plastic scintillator target~\cite{MINERvA:2019wqe,MINERvA:2023ikp,MINERvA:2023avz,ref:LastThesis}. MINERvA has measured multiplicity and production cross sections and has had some success reconstructing the kinematics for higher kinetic energy neutrons in antineutrino interactions.  

The SuperFGD is the component of the recently completed T2K ND280 upgrade~\cite{T2K:2019bbb} that is  based on 1~cm scintillator cubes read out along three orthogonal dimensions.  It can detect neutrons via isolated energy depositions and determine \Tn via Time-of-Flight (ToF)~\cite{Munteanu:2019llq}.
Results using this technique on data are expected soon.
Measurements from this detector will be useful for constraining the interaction and beam models currently used in T2K and in the future for the Hyper-Kamiokande~\cite{Hyper-Kamiokande:2018ofw} experiment.
These data are also exciting because the detector should have the timing and spatial resolution to enable the selection of a sample enriched in antineutrino interactions on the hydrogen in the scintillator target.

Recent work by MicroBooNE demonstrates the reconstruction of protons knocked out by neutrons in a liquid argon detector, effectively tagging the presence of the neutron in the event~\cite{MicroBooNE:2024hun}.  
This work is important, in part, because the interactions are on argon in a liquid argon TPC.  This means that DUNE will not be completely blind to neutrons in far detector interactions or in near detector data used to constrain the beam and interaction models that facilitate the extraction of oscillation parameters from measurements in the far detector.
A shortcoming of the MicroBooNE work is that the technique has little sensitivity to \Tn. 

One aspect of the MicroBooNE technique is that the direction of the neutron is well determined by the line joining the neutrino interaction vertex and the initial point of the proton track in the limit that the neutron has not undergone some significant undetected scattering along the way.  In T2K, the direction of the neutron that deposits energy is also determined well in the same limit.  
This paper explores the usefulness of this direction in inferring \Tn, improving the neutrino energy reconstruction, and improving the neutron tag efficiency and/or purity.  

\section{Inferring Neutron Kinetic Energy Using Direction}
\label{section:technique}

\begin{figure*}  
	\centering	\includegraphics[width=0.8\textwidth]{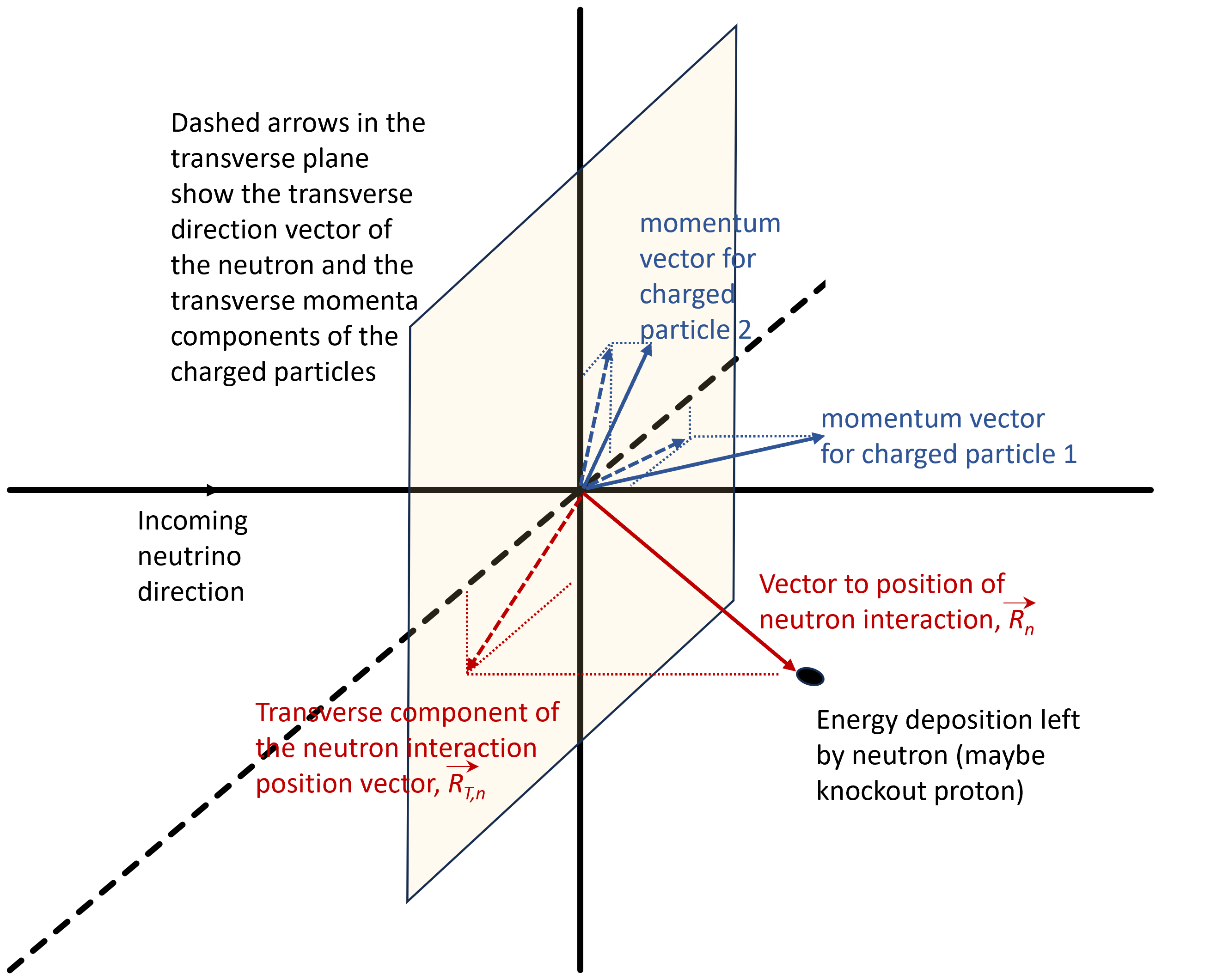}

	\caption{Sketch of a neutrino interaction with the momenta of the reconstructed charged particles and the direction of the produced neutron projected onto the transverse plane of the interaction. }
	\label{fig:neutrino_event}
\end{figure*}

Figure~\ref{fig:neutrino_event} shows a schematic of a generic neutrino charged-current interaction.  The momenta of reconstructed particles, such as muons, protons, and both charged and neutral pions, are shown as blue vectors.
Also shown in Fig.~\ref{fig:neutrino_event} is the direction of a neutron as tagged by some sort of neutron interaction leaving an energy deposition in the detector.
In the case of MicroBooNE~\cite{MicroBooNE:2024hun}, the energy deposition is due to proton knockout and the characteristics of the proton track are useful for reducing backgrounds. For this discussion, the specificity of the proton knockout interaction is not necessary. 
Regardless of the interaction that creates the energy deposition giving the neutron tag, the neutron direction is assumed to be in the same direction relative to the neutrino interaction vertex as the neutron tag energy deposition.
This ignores the practical, experimental need to have neutron tags with characteristics that are distinguishable from backgrounds and sources of noise that mimic neutron signals.

\begin{figure*}  
	\centering	\includegraphics[width=0.9\textwidth]{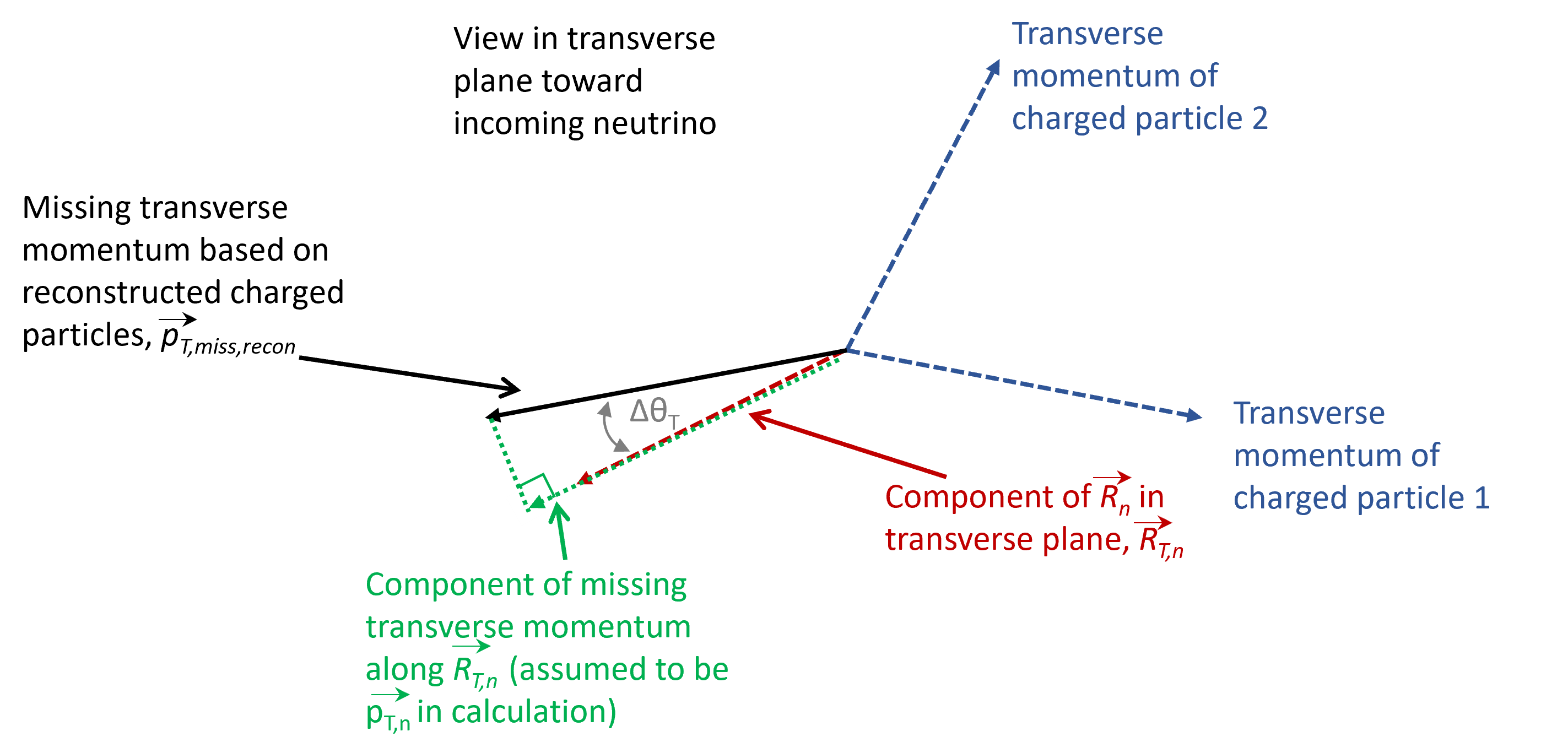}

	\caption{View of transverse plane of the interaction toward the incoming neutrino direction showing the reconstructed charged particle transverse momentum vectors, the missing transverse momentum vector and the neutron direction projected onto the transverse plane.}
	\label{fig:transverse_plane}
\end{figure*}

In Fig.~\ref{fig:transverse_plane}, the momentum vectors of the reconstructed particles and the direction vector of the neutron are projected onto the plane transverse to the neutrino direction which is well known for neutrinos produced by accelerators.
With perfect reconstruction of all produced particles and nuclear recoil, there will be momentum conservation in the transverse plane and the total transverse momentum should sum to zero.
Perfect reconstruction does not exist.
Still, the limit of transverse momentum conservation is useful even in the world of imperfect detectors.  With the assumption of transverse momentum conservation for the reconstructed particles, the transverse momentum sum can be used to infer the ``missing'' transverse momentum, \ptmr.
This vector is also shown in Fig.~\ref{fig:transverse_plane}.

The vector \ptmr does not contain contributions from the tagged neutron and other unreconstructed particles and
it is smeared out by nuclear effects, including recoil.
That said, in the limits of small nuclear effects and no significant missing momentum from unreconstructed particles other than the tagged neutron,  \ptmr will line up along the projection of the position vector for the neutron interaction, \Rn, in the transverse plane, \Rtn.
By construction, the opening angle between \ptmr and \Rtn which is defined here as \dt, will become larger in general as the nuclear effects, missing momentum of the unreconstructed particles, and recoil increase relative to the momentum of the tagged neutron.  

Under the assumption that $\ptmr \cos{\dt}$, the component of \ptmr parallel to \Rtn, is equal to the magnitude of the transverse momentum of the tagged neutron, \ptn, the momentum of the neutron \pn can be determined using
\begin{equation}
    \label{eqn:sim}
    |\pn| = |\ptn|\frac{|\Rn|}{|\Rtn|}.
\end{equation}
Using Eq.~\ref{eqn:sim} as well as the known position of the tagged neutron interaction, \Rn, the kinetic energy of the neutron, \Tn, can be calculated and all the information needed to include the neutron in the event reconstruction is known. 

The utility of this technique depends on the validity of the assumptions of no nuclear effects and no significant missing momentum from unreconstructed particles other than the tagged neutron.
Note that \dt might be a useful diagnostic for flagging events where the assumptions are bad, i.e., events with large \dt.
That angle may also prove to be useful for tagging events in which the neutrino interaction occurred on a hydrogen nucleus in a hydrocarbon target.
Such events would show up at small \dt.

This technique has some similarity to one proposed by G.~Myatt over 50 years ago to infer the energy of neutral particles escaping from neutrino interactions in hydrogen bubble chambers~\cite{myatt}.  In Ref.~\cite{myatt}, the assumption is made that the missing particle goes in the direction of the total momentum of the reconstructed hadrons, which tends to be more true at the high neutrino energies Myatt considered than what is being considered in this work.  Here, the energy is lower, and the direction of the neutron is known.

\section{Initial assessment of the technique through simulations}

\subsection{Description of study done using simulated events}

To determine the potential usefulness of this technique, a study was performed using the GENIE neutrino event generator~\cite{genie0}, version 3.04.02, along with tune \texttt{G18\_02a\_00\_000}.
The NUISANCE software framework~\cite{nuis:2016jfr}, version 2.9.9, was used to convert the GHEP output from GENIE into flat root trees for analysis~\footnote{The version of GENIE and NUISANCE used is pre-compiled inside a container made by L.~Pickering and can be found at \url{https://hub.docker.com/r/picker24/nuisance2_git_master/tags}}.

The study was carried out at the generator level making use of truth information along with nominal detection thresholds and smearing.
All events in the study were required to be $\nu_{\mu}$ or $\bar{\nu}_{\mu}$ charged-current (CC) interactions.
Charged particles and neutral pions coming from the interaction were considered.
The neutral pions were assumed to decay into photons which, in turn, interact and produce detectable particles.
Produced neutrons, other than the one considered to be tagged, were ignored as undetected particles.



The aim of this study was to determine the potential usefulness of this reconstruction technique in a liquid argon TPC-based detector such as DUNE, or a 3D scintillator detector such as the SuperFGD in the T2K near detector.  
Therefore, the detection thresholds and smearing applied to the detectable particles in the study were chosen to be something plausibly used for these detectors.
These thresholds and smears are summarized in~\autoref{tab:vis_smear_thresh}.
The momentum smears were assumed Gaussian in magnitude, and the direction was smeared Gaussian in solid angle.
For the T2K-like analysis, the angular smear is in the middle of the expected range for charged tracks in the SuperFGD.
For the DUNE-like analysis, the directional smearing is small enough that it gave little effect in the results and was ignored.
The detection thresholds were applied to the smeared quantities.

Neutrino and antineutrino interactions on argon were studied.  For these events, the neutrino(antineutrino) flux was assumed to be roughly that expected for the forward(reverse)  horn current (FHC and RHC, respectively) LBNF beam at the near detector~\cite{DUNE:2021cuw}. Similarly, neutrino(antineutrino) interactions on carbon, hydrogen, and scintillator (CH) were studied where the neutrinos(antineutrinos) had energy spectra similar to those seen in the T2K off-axis FHC(RHC) neutrino beam at ND280~\cite{nd280flux}. 

\begin{table*}[hbt]
    \centering
    \begin{NiceTabular}{c | c | c | c}[hvlines]
    \hline
        {\bf Target nucleus} & \textbf{Reco. Particle} & \textbf{Smearing $\bm{|p|}$, direction}  & \textbf{Threshold (\mev/c)}\\ 
    \hline\hline
   \Block{4-1}{C, CH}& proton, $p$        &  5\%, 5$^{\circ}$ &    335  \\ 
    \cline{2-4}
      & charged pion, $\pi^\pm$      &  20\%, 5$^{\circ}$ &   128 \\ 
    \cline{2-4}
     & neutral pion, $\pi^0$      &  20\%, 5$^{\circ}$ &   0  \\ 
    \cline{2-4}
     & muon, $\mu$ & 3\%, 5$^{\circ}$ &  85.1 \\
\hline     
    \Block{4-1}{Ar}& proton, $p$        &  6\%, none &    301  \\ 
    \cline{2-4}
      & charged pion, $\pi^\pm$      &  20\%, none &   96.3 \\ 
    \cline{2-4}
     & neutral pion, $\pi^0$      &  12\%, none &   0  \\ 
    \cline{2-4}
     & muon, $\mu$ & 3\%, none &  85.1 \\
    \cline{2-4}
    \hline\hline
    \end{NiceTabular}
    \caption{ Particles deemed reconstructible for this study along with the parameters representing reconstruction thresholds and detector smearing effects. }
    \label{tab:vis_smear_thresh}
\end{table*}

Potential neutron candidates were required to have $\Tn>20$~\mev so that they could leave, in principle, a plausibly detectable energy deposition.
For this study, if one or more neutrons were produced, the one with the highest kinetic energy was chosen as the neutron candidate for reconstruction.
Only one neutron candidate was chosen per neutrino interaction. The efficiency of detection and backgrounds were not considered.
It was assumed that the neutron of interest was tagged via an energy deposition in SuperFGD cubes or liquid argon.
Also, it was assumed that the spatial vector from the neutrino interaction point to the tagged energy deposition gave a good measure of the initial direction of the neutron.
This quantity was not smeared.  

\subsection{Interactions on C and Ar}

\begin{figure*}  
	\centering	\includegraphics[width=0.45\textwidth]{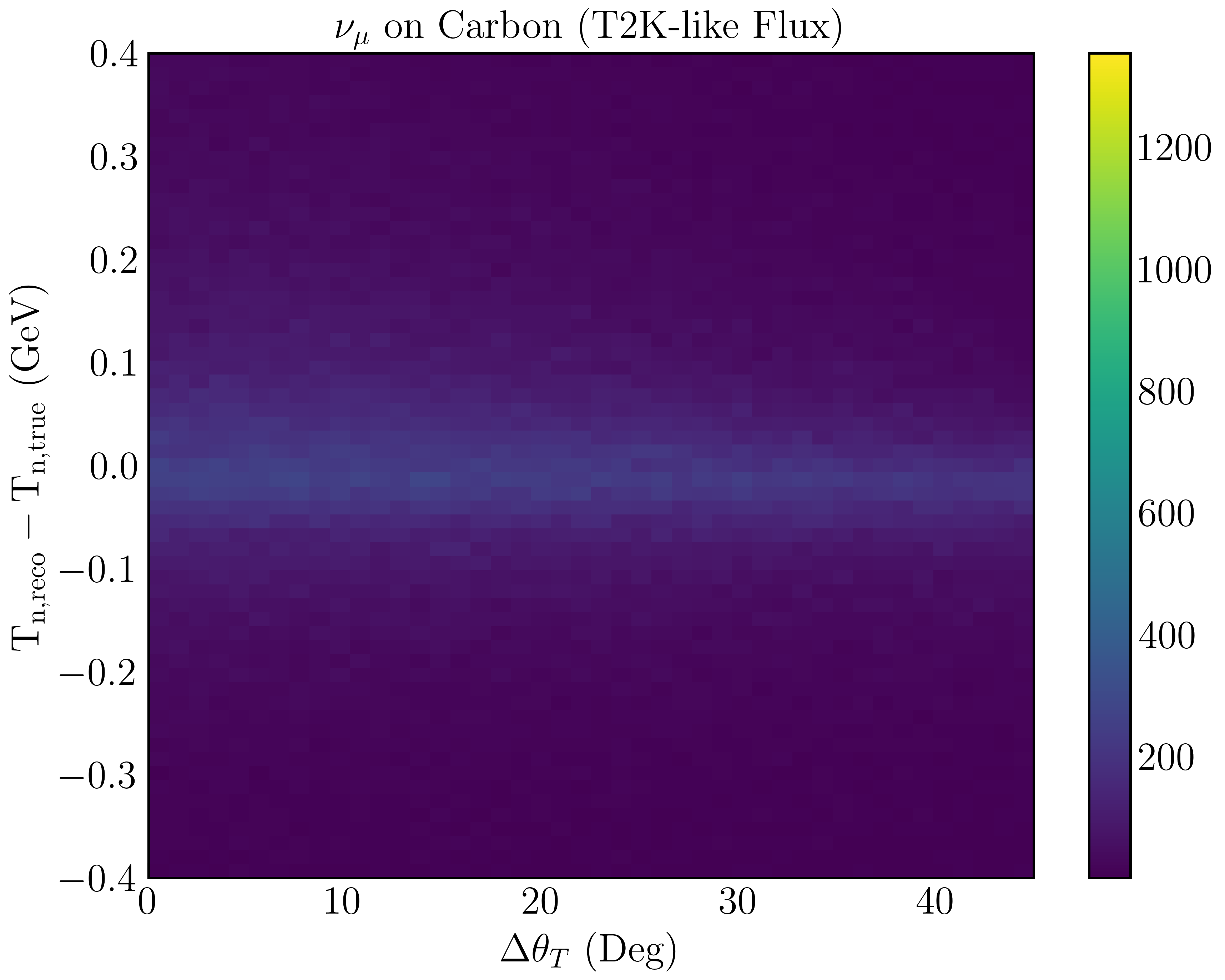}
    \includegraphics[width=0.45\textwidth]{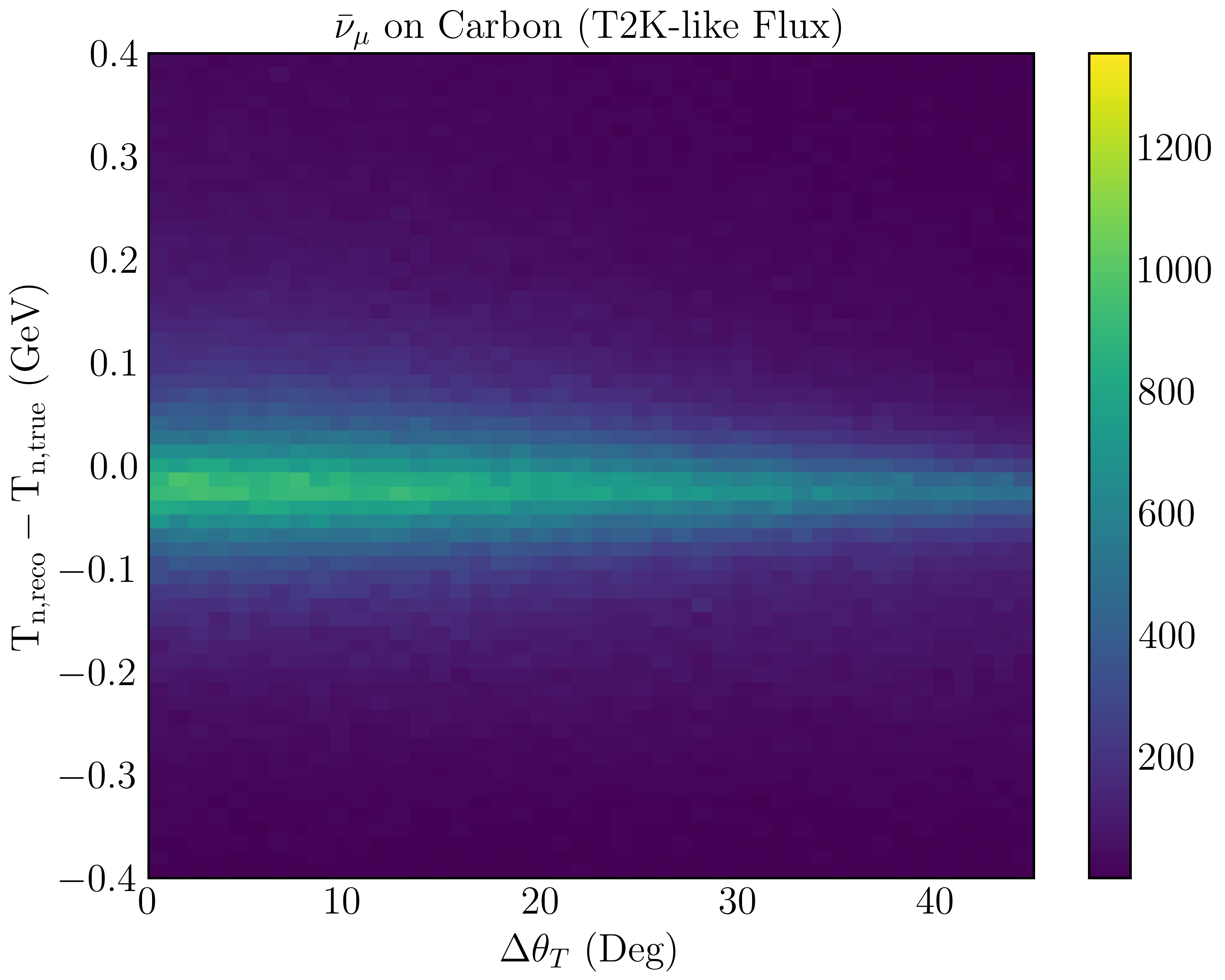}
    \includegraphics[width=0.45\textwidth]{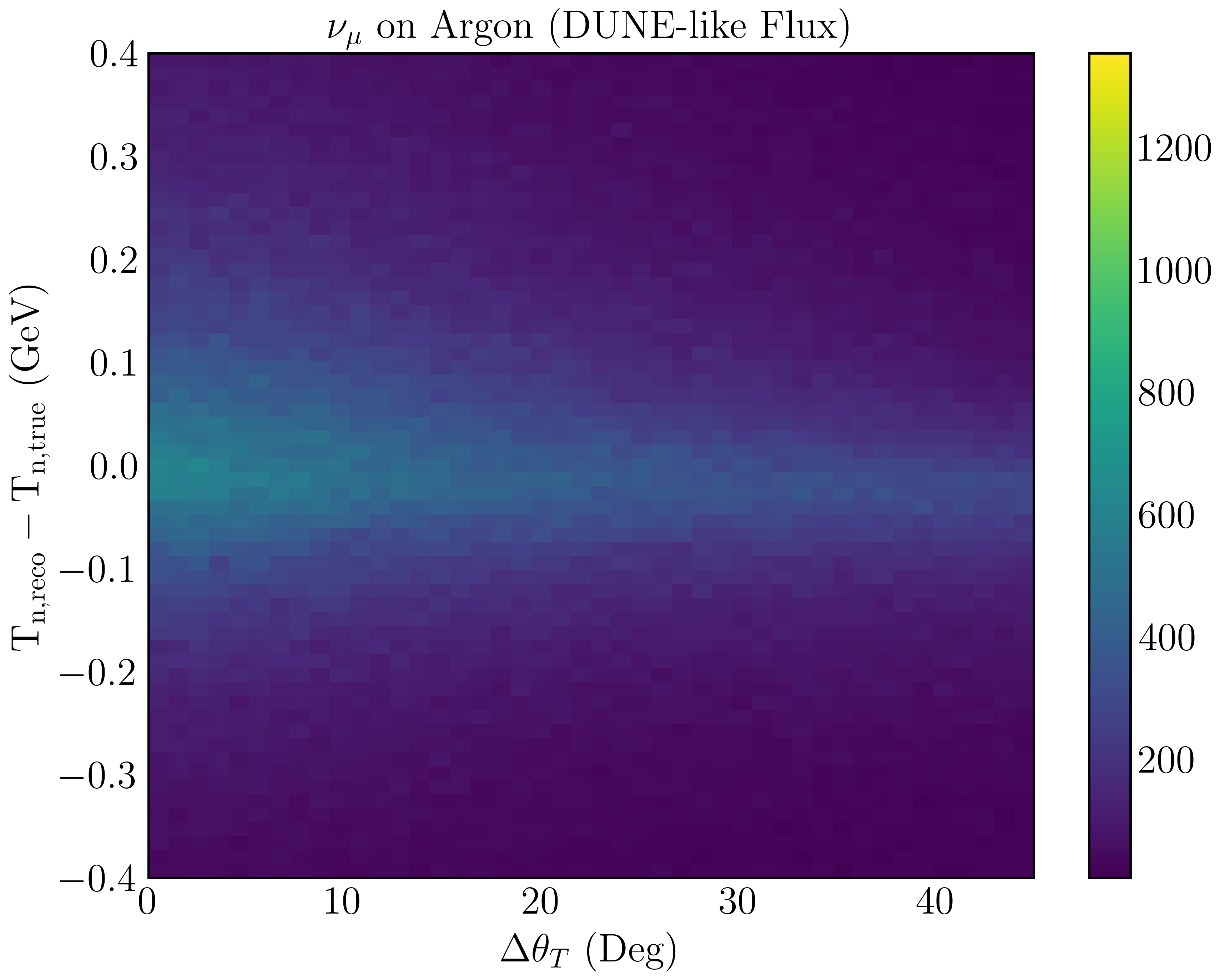}
    \includegraphics[width=0.45\textwidth]{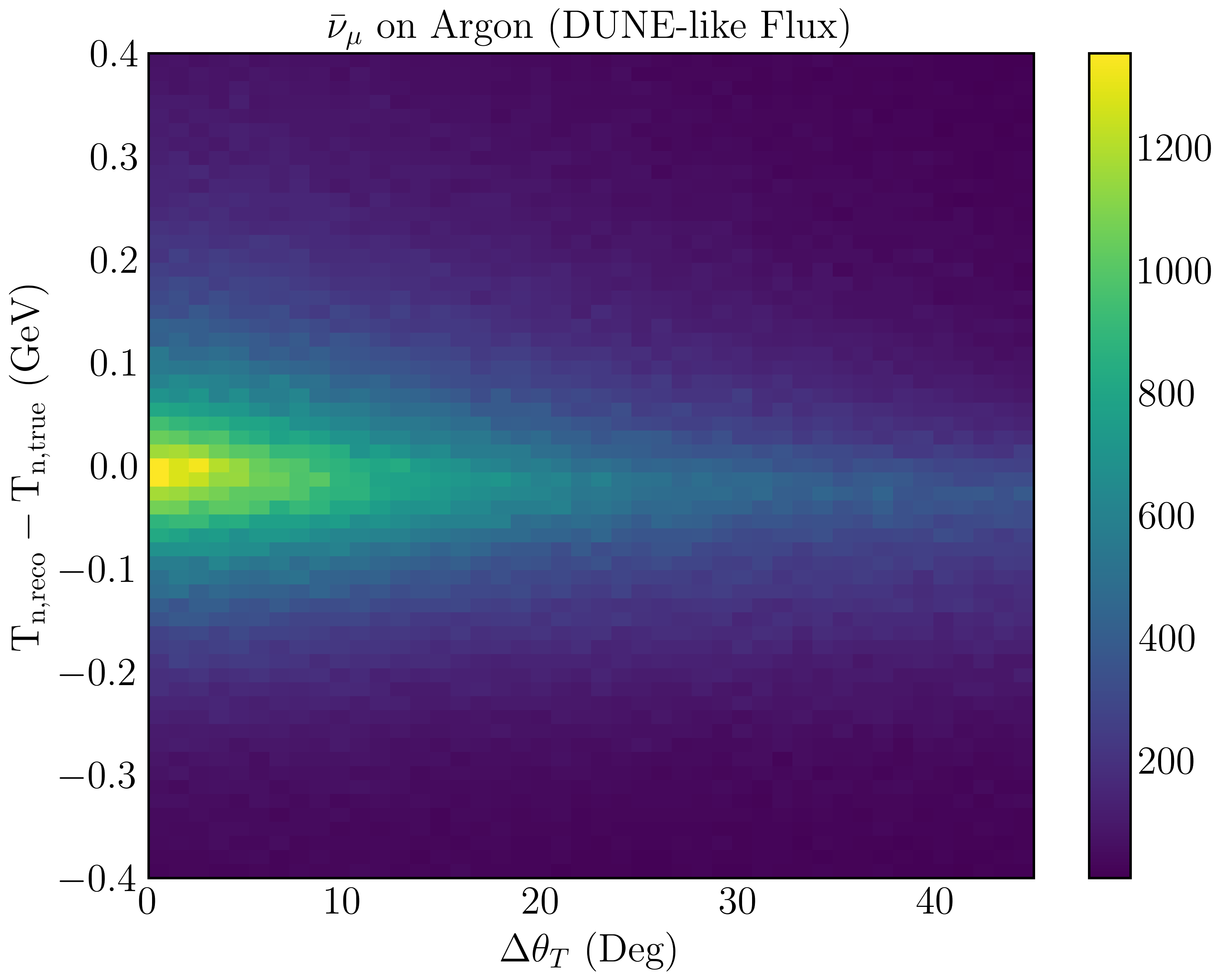}  
	\caption{These plots show \dt versus $\Tnr-\Tnt$ for the selected neutron candidate.  The top two plots show interactions on carbon for $\nu_{\mu}$ (left) and $\bar{\nu}_{\mu}$ interactions, respectively, in a T2K-like flux.  On the bottom are the analogous plots for interactions on argon in a DUNE-like flux.  }
	\label{fig:dt_ke}
\end{figure*}

 Using the technique outlined in Sec.~\ref{section:technique}, Figure~\ref{fig:dt_ke} shows \dt versus the difference between the reconstructed neutron kinetic energy, \Tnr, and the true neutron kinetic energy, \Tnt, according to the simulation for all selected neutron candidates. 
 The upper left panel is $\nu_{\mu}$ interactions on a carbon nucleus, assuming an FHC T2K-like flux.
 The upper right panel shows $\bar{\nu}_{\mu}$ interactions on carbon using an RHC T2K-like flux. 
 The bottom shows results for $\nu_{\mu}$ and $\bar{\nu}_{\mu}$ interactions on an argon nucleus assuming FHC and RHC DUNE-like fluxes, respectively.
 Note that the bulk of the events show up at low \dt reflecting the expected correlation between \Rtn and \ptmr.
 The differences between the $\nu_{\mu}$ and $\bar{\nu}_{\mu}$ distributions are thought to be driven by the different hadronic energy transfer (y) distributions and neutron production mechanisms.  
 

 Many background sources, such as noise or neutrons arising from other interactions or out-of-fiducial volume events, should be distributed flat in \dt.
 This suggests that a cut at low \dt might be useful in enhancing the signal to background in a real analysis. For this study, $\dt < 30^\circ$ was chosen since it was thought to be small enough to help reduce those events where the correlation in $\dt$ was lost due to something strange happening in the event.  The cut was also thought to be large enough that the cut performance would not have much sensitivity to the particle reconstruction thresholds.  A more aggressive cut or careful optimization was not warranted given the exploratory nature of this study.  The $\dt < 30^\circ$ cut removed approximately half the events in each of the samples considered.

\begin{figure*}  
	\centering	\includegraphics[width=0.45\textwidth]{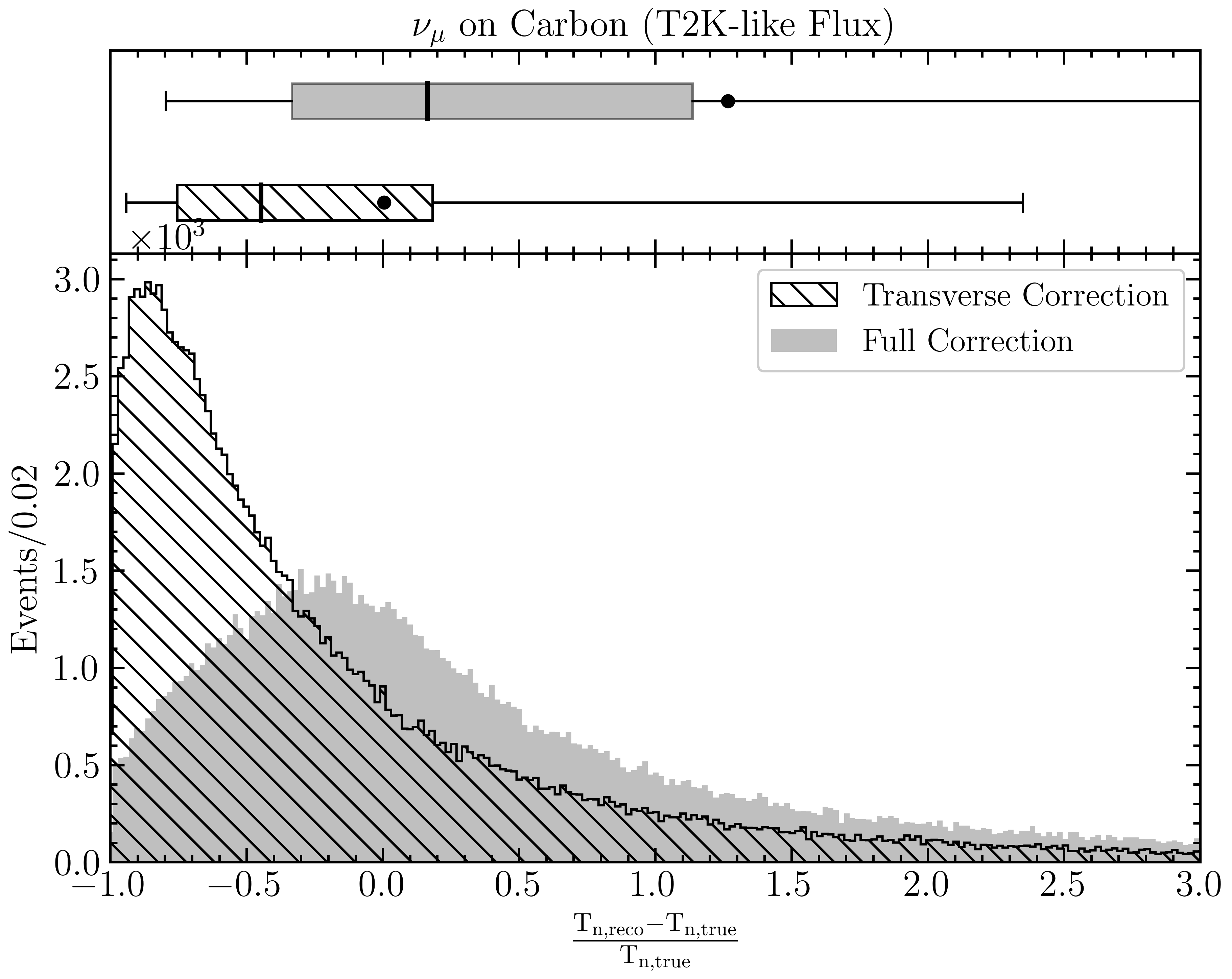}
    \includegraphics[width=0.45\textwidth]{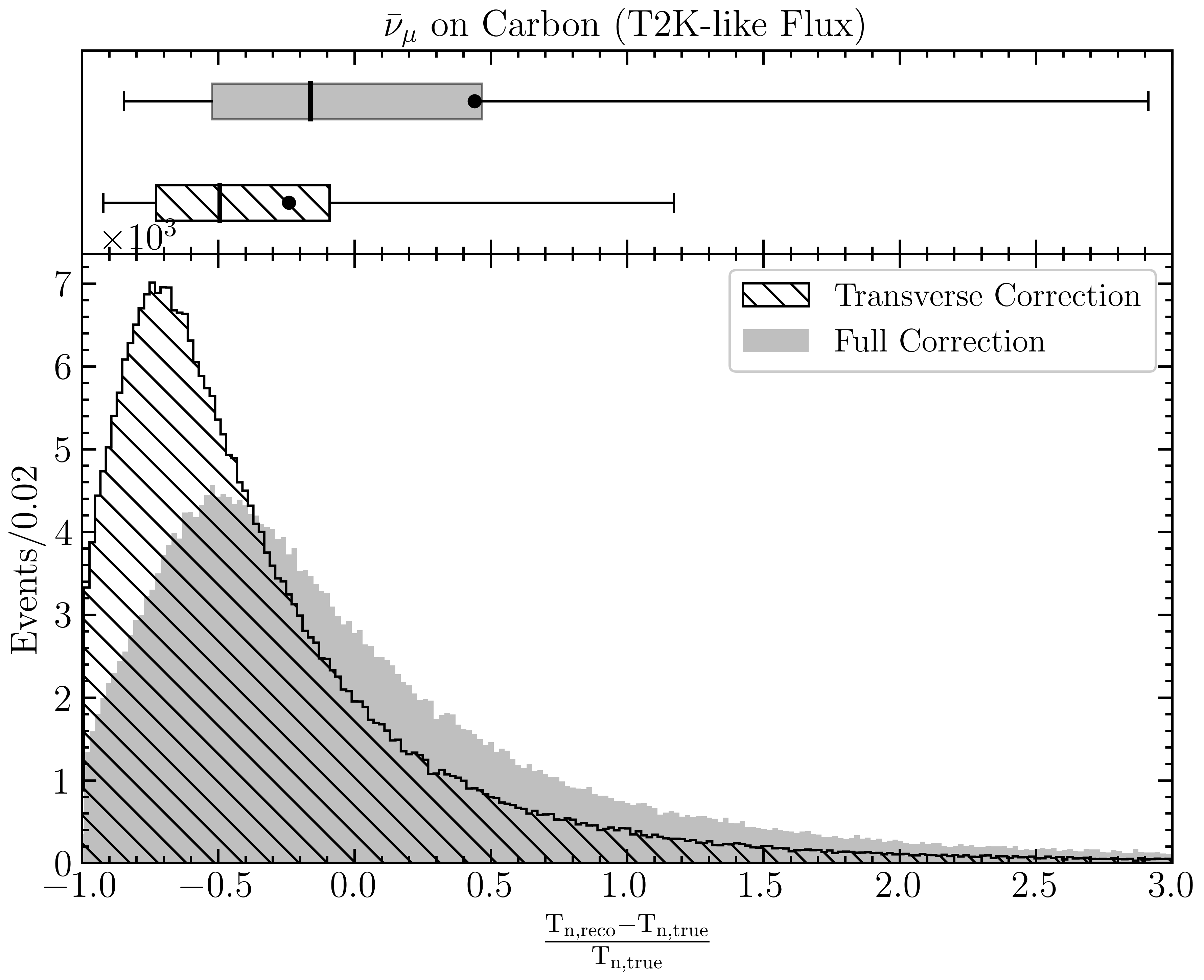}
    \includegraphics[width=0.45\textwidth]{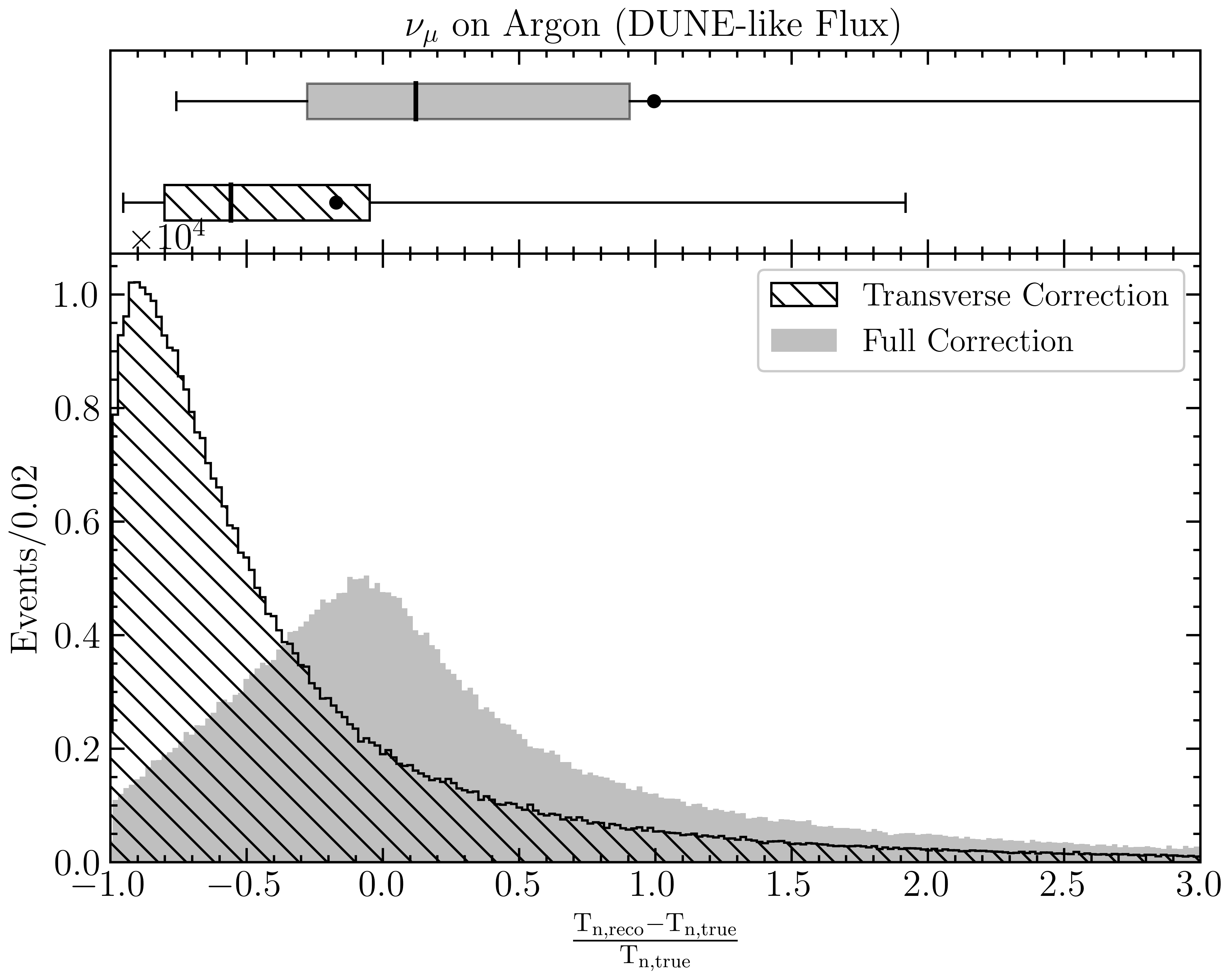}
    \includegraphics[width=0.45\textwidth]{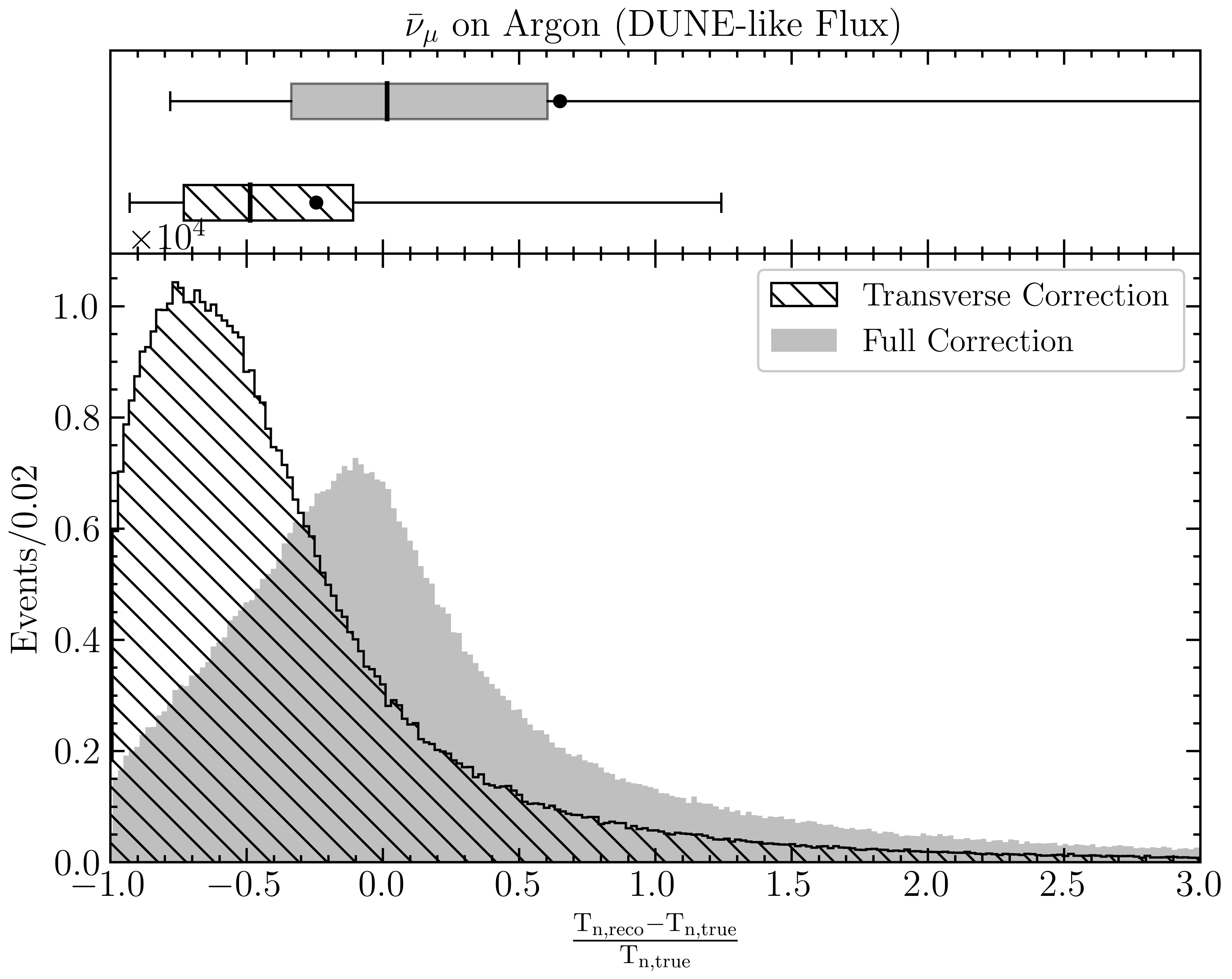}
	\caption{Reconstructed neutron kinetic energy, \Tn, resolution for the selected neutron candidates using the technique presented in  Sec.~\ref{section:technique} as the solid histogram.  The top two plots show interactions on carbon for $\nu_{\mu}$ (left) and $\bar{\nu}_{\mu}$  (right) interactions, respectively, in a T2K-like flux.  On the bottom are the analogous plots for interactions on argon in a DUNE-like flux. The hatched histogram shows the same thing where the  \Tn is derived from assuming the magnitude of the missing transverse momentum is the magnitude of the neutron momentum. The box and whiskers plot at the top of each panel shows the mean of each distribution as a solid black dot and the median as a solid black vertical line.  The central 50\% of the events lie within the box and the whiskers show the central 90\% of the events. }
	\label{fig:n_ke_res}
\end{figure*}

After implementing a $\dt < 30^\circ$ cut and correcting for the mean kinetic energy bias using the simulation, the neutron candidate kinetic energy resolution determined using the technique presented in Sec.~\ref{section:technique} is shown in Fig.~\ref{fig:n_ke_res} in the solid gray histogram. The hatched histogram represents a related method that will be discussed later.  The mean kinetic energy bias mentioned above is 3~MeV($-$21~MeV) and 8~MeV($-$11~MeV) for (anti)neutrinos on C and Ar, respectively.
The breakdown of the panels in terms of target type and incident (anti)neutrino is the same as shown in Fig.~\ref{fig:dt_ke}.
The distributions are very non-Gaussian and biased toward an underestimate of \Tn.
The Ar distributions are narrower and more centered than those for interactions on C. 
To understand this better, additional simulations were performed where the target nuclei were swapped, i.e., done using the Ar target with the T2K-like flux and directional smearing and the C target with the DUNE-like flux and smearing.  These simulations showed that the differences between the C and Ar results in Fig.~\ref{fig:dt_ke} are driven by the different (anti)neutrino spectra and the directional smearing applied to the reconstructed particles and not so much the variation in target mass.

To further characterize these results, the upper panel of each plot is a ``box and whiskers'' distribution, where the mean is shown as the black circle and the median is shown as the dark vertical line.
The central 50\% of the events are enclosed by the box.
Said a different way, the sides of the box are drawn at the quartile point from the median in each direction.
The whiskers show the central 90\% of the events.
The difference between the median and the mean is driven by the asymmetry of the distribution and the size of the tails on each side.  

In these and many other plots shown in this paper, the distributions have significant tails and asymmetries.
This means that any discussion of the width of the distribution in terms of a Gaussian-like ``$\sigma$'' or half-width at half-max is problematic.
Nevertheless, a metric characterizing the width will be useful for the discussion.
The metric used for this will be called $\sigma'$ and is defined as the half-width of the length that would encompass the central 68\% of the events in the distribution centered on the median by taking 34\% to each side of the median.
In most of these plots, the length from which $\sigma'$ is calculated is not symmetric about the median.
For a normal distribution, the mean is the same as the median and $\sigma'$ would correspond to the $\sigma$ of the normal distribution.

For comparison, also shown in Fig.~\ref{fig:n_ke_res}, in the hatched histogram, is the \Tn resolution assuming the magnitude of the missing transverse momentum is equal to the total neutron momentum.  This is a technique that is easy to apply without directional information for the neutron(s) produced in the event.  In the figure it can be seen that the technique presented in this paper that makes use of the neutron direction gives an asymmetric resolution in \Tn more centered at zero than the other technique, i.e., closer to the true \Tn. Both techniques have large tails.

\begin{figure*}  
	\centering	\includegraphics[width=0.45\textwidth]{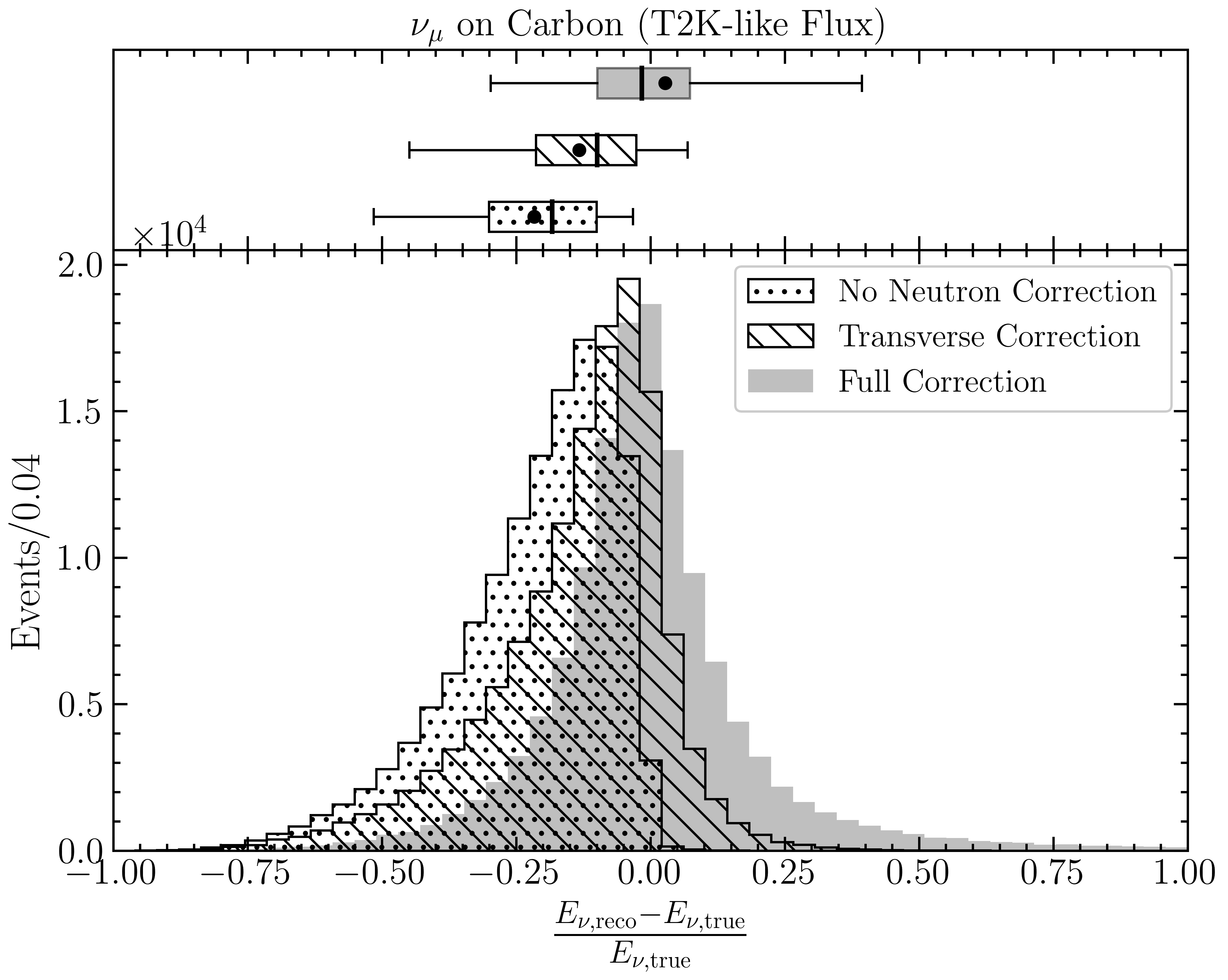}
    \includegraphics[width=0.45\textwidth]{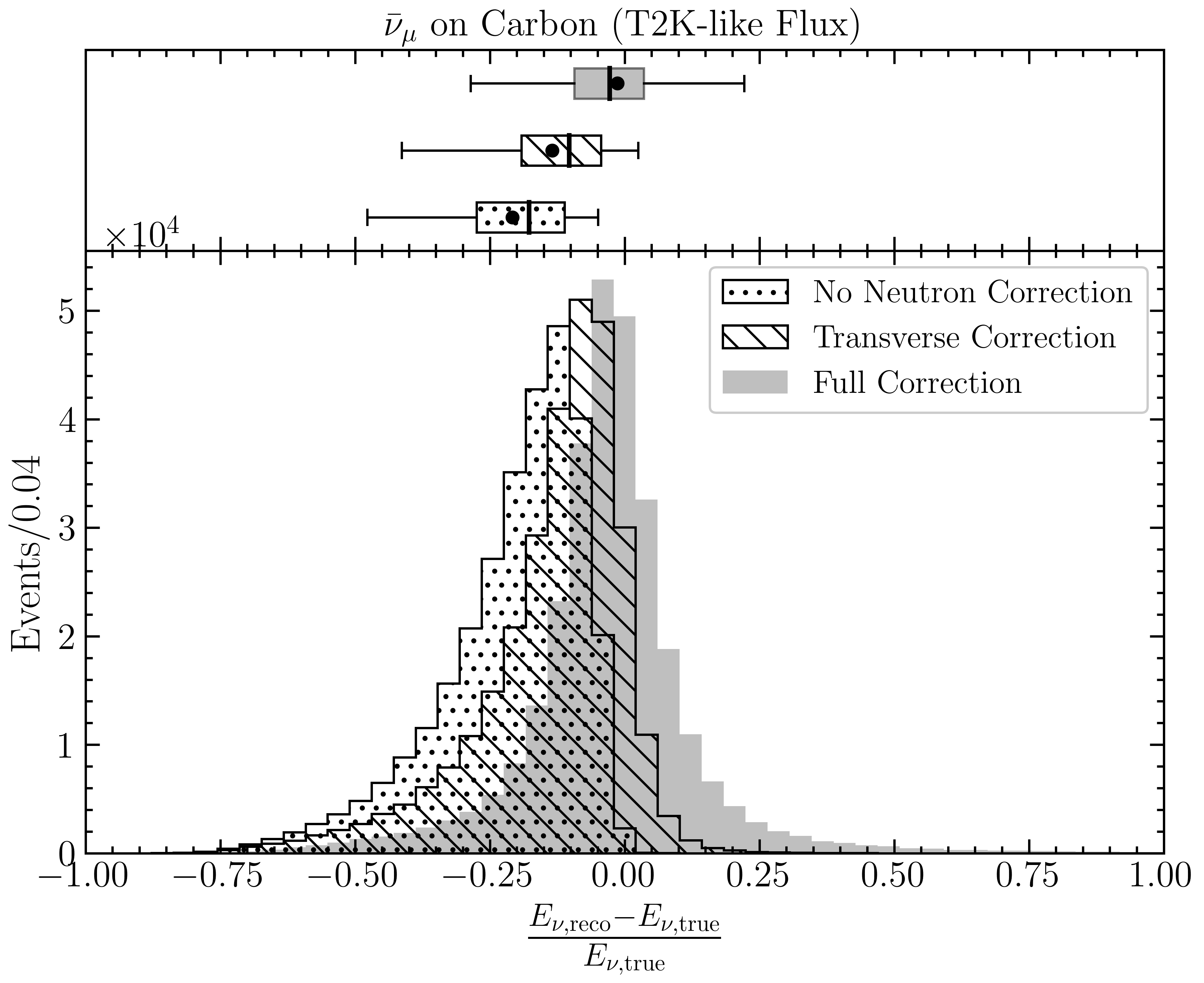}
    \includegraphics[width=0.45\textwidth]{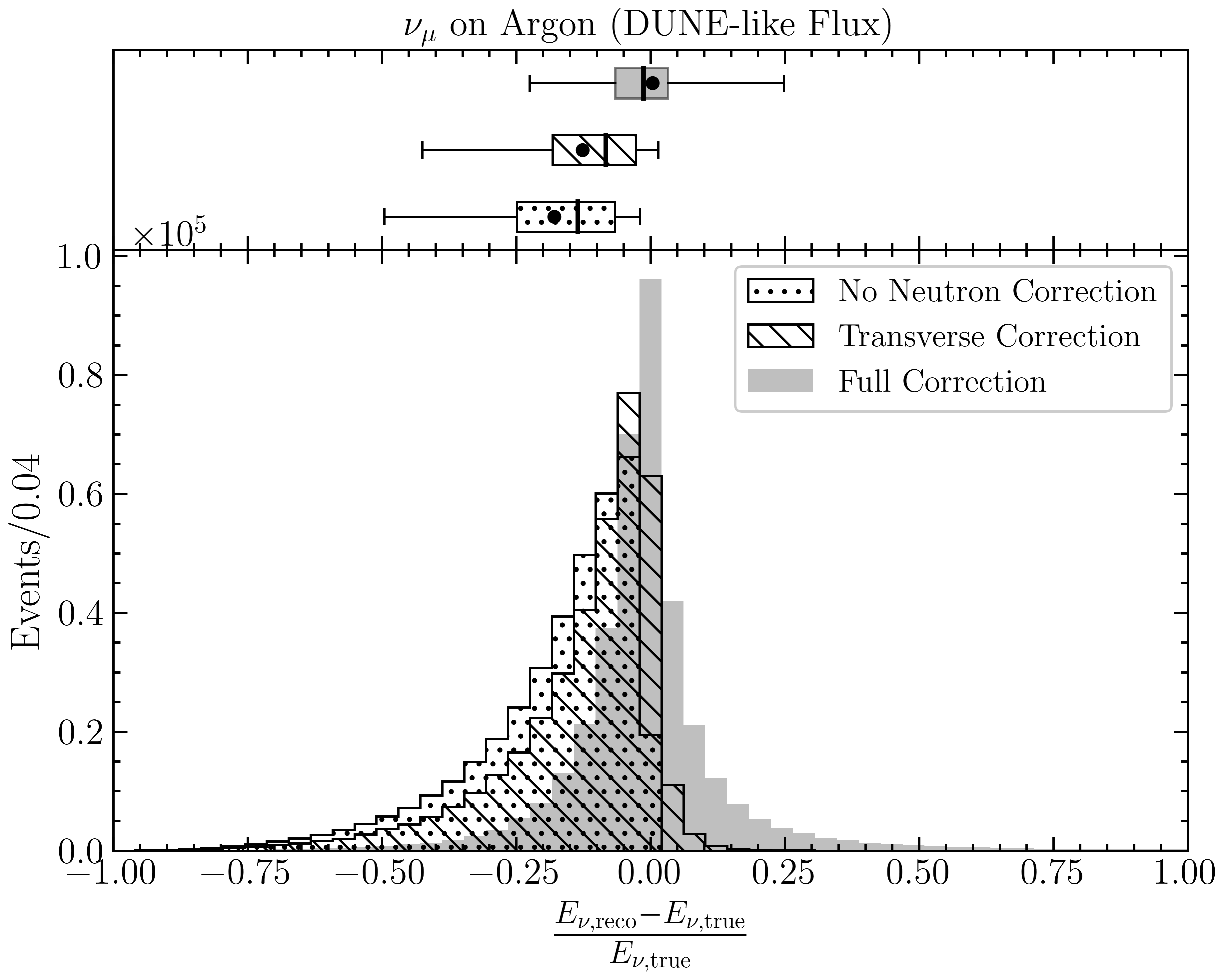}
    \includegraphics[width=0.45\textwidth]{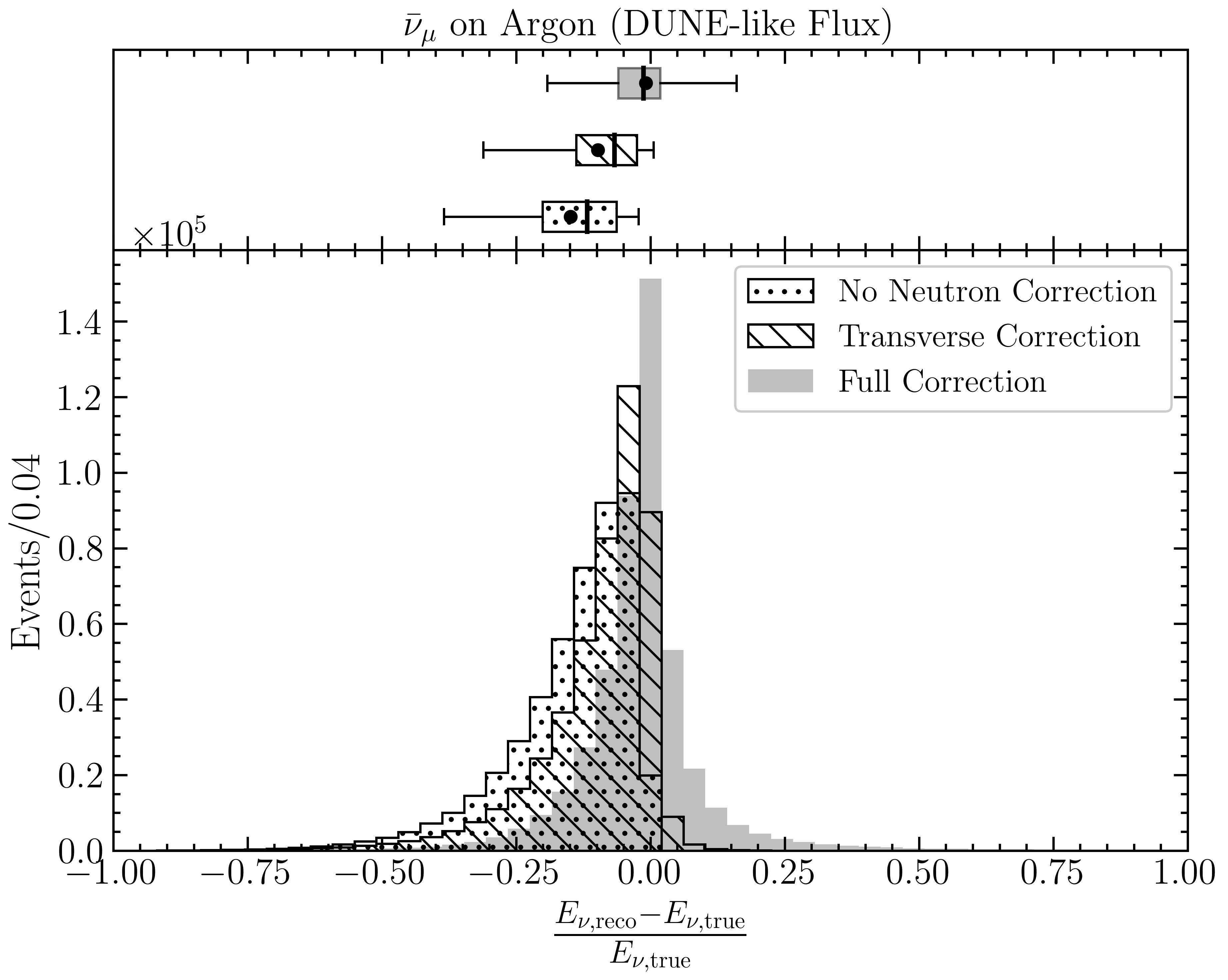}
	\caption{Reconstructed (anti)neutrino energy resolution for the selected events.  The top two plots show interactions on carbon for $\nu_{\mu}$ (left) and $\bar{\nu}_{\mu}$ interactions, respectively, in a T2K-like flux.  On the bottom are the analogous plots for interactions on argon in a DUNE-like flux. The solid histogram shows the energy resolution calculated using \Tn for the selected neutron candidates as determined by the technique presented in  Sec.~\ref{section:technique} (full correction).  The hatched histogram assumes \Tn to be determined from the missing transverse momentum alone (transverse correction).   The dotted histogram assumes all the energy from neutrons is missing in the calculation (no neutron correction). The box and whiskers plot at the top of each panel shows the mean of each distribution as a solid black dot and the median as a solid black vertical line.  The central 50\% of the events lie within the box and the whiskers show the central 90\% of the events.}
	\label{fig:enu_res}
\end{figure*}

It is important to examine how the inclusion of the reconstructed neutron affects the reconstructed (anti)neutrino energy resolution.  Fig.~\ref{fig:enu_res} shows the (anti)neutrino energy resolution for three different algorithms.  One algorithm, shown as the histogram with a dotted pattern in the panels of the figure, ignores the produced neutrons and determines the neutrino energy via a calorimetric sum of the observed energy.   The second algorithm, shown in the lined hatched histogram in the panels of the figure, adds in the kinetic energy of the produced neutron(s) as determined by setting the magnitude of the neutron momentum equal to the magnitude of the missing transverse momentum.  The third technique, shown as the solid histogram, adds in the neutron kinetic energy as determined using the technique presented in Sec.~\ref{section:technique}. The breakdown of the panels in terms of target type and incident (anti)neutrino is the same as shown in the previous figures.

The first two energy-reconstruction algorithms significantly underestimate the neutrino energy.
The third algorithm gives a slightly asymmetric distribution that is somewhat Gaussian in shape and centered on the true neutrino energy.
The lower-energy tail is more pronounced than the higher-energy tail.   
As shown in Tab.~\ref{tab:widths}, the width measure, $\sigma'$, for the reconstruction technique presented in Sec.~\ref{section:technique} is  15\%/11\%/9\%/7\% for $\nu_{\mu}$  on C / $\bar{\nu}_{\mu}$ on C / $\nu_{\mu}$ on Ar / $\bar{\nu}_{\mu}$ on Ar, respectively.
Given the fairly poor integrated \Tn resolution yielded by the technique, it is a little surprising that the neutrino energy resolution is so good.  Investigation shows that the tail in the \Tn resolution is populated mostly by lower \Tn candidates that are produced at a large angle with respect to the neutrino direction, i.e., more transverse.  The higher \Tn neutrons tend to be produced at smaller angle with respect to the neutrino.  This is a region where the longitudinal momentum is larger and the correction made by the scheme presented in this paper is larger.  The \Tn resolution is much better for the higher \Tn candidates.  The \Tn resolution is decent where it matters most and poor where it makes less difference to the overall neutrino energy reconstruction.    

\begin{table*}[hbt]
    \centering
    \begin{NiceTabular}{c | c | c | c | c | c}[hvlines]
    \hline
        {\bf Target nucleus} & \textbf{$\bm{\nu}$ type} & \textbf{Quantity}  & {$\bm{\sigma'}_\textbf{No Correction}$} & {$\bm{\sigma'}_\textbf{Transverse Correction}$} & {$\bm{\sigma'}_\textbf{Full Correction}$}  \\ 
    \hline\hline 
   \Block{4-1}{C} & \Block{2-1}{$\nu_{\mu}$}  
    & E$_{\nu}$ & 0.148 & 0.143 & 0.145 \\
    & & KE$_{n}$ & --- & 0.762 & 1.274 \\
    & \Block{2-1}{$\bar{\nu}_{\mu}$} & E$_{\nu}$ & 0.124 & 0.116 & 0.107 \\
    & & KE$_{n}$ & --- & 0.508 & 0.811 \\
    \hline\hline
   \Block{4-1}{Ar} & \Block{2-1}{$\nu_{\mu}$}
    & E$_{\nu}$ & 0.139 & 0.119 & 0.089 \\
    & & KE$_{n}$ & --- & 0.632 & 1.056 \\
    & \Block{2-1}{$\bar{\nu}_{\mu}$} & E$_{\nu}$ & 0.104 & 0.086 & 0.070 \\
    & & KE$_{n}$ & --- & 0.502 & 0.834 \\  
    \hline\hline
    \end{NiceTabular}
    \caption{The distribution width metric, $\sigma'$, for the \Tn resolution and neutrino energy resolution plots shown in Fig.~\ref{fig:n_ke_res} and Fig.~\ref{fig:enu_res}, respectively.  The column labeled ``No correction'' gives the values for the case where neutrons are not included in the calculation.   The column labeled ``Transverse correction'' gives numbers where the quantities are calculated assuming the transverse momentum is the neutron momentum.   The column labeled ``Full correction'' gives the numbers where \Tn is calculated by the technique presented in  Sec.~\ref{section:technique}.}
    \label{tab:widths}
\end{table*}

\begin{figure*}  
	\centering	\includegraphics[width=0.60\textwidth]{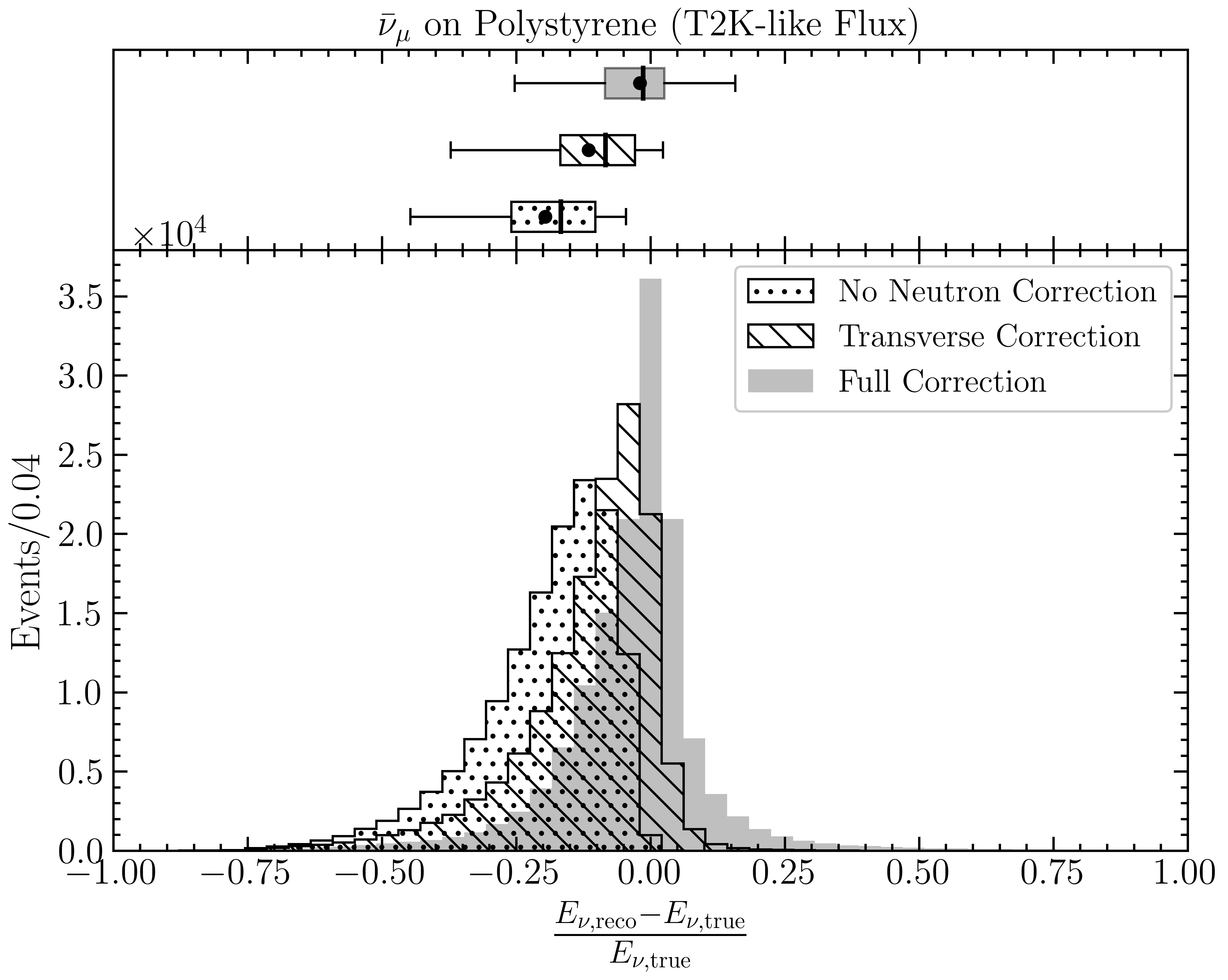}
	\caption{The energy resolution for antineutrino interactions on CH in a T2K-like flux for $\dt < 10^\circ$. The different histograms show the energy resolution calculated three different ways as described in the caption for Fig.~\ref{fig:enu_res}.  The box and whiskers plot at the top of the panel shows the mean of the distribution as a solid black dot and the median as a solid black vertical line.  The central 50\% of the events lie within the box and the whiskers show the central 90\% of the events.}
	\label{fig:enu_res_CH}
\end{figure*}

A similar neutron energy reconstruction resolution plot for $\bar{\nu}_{\mu}$ on polystyrene scintillator (CH) for a T2K-like flux is shown in Fig.~\ref{fig:enu_res_CH} for events satisfying a cut demanding $\dt<10^{\circ}$.  
Here, the distribution based on the technique presented in Sec.~\ref{section:technique} is narrower than that on pure C due to the inclusion of events where the neutrino interacts on H.  In those interactions on H the reconstruction works quite well since there are no nuclear FSI or recoil effects.
In this case, $\sigma'$ is 9.6\%.
A selection based on a tight cut in \dt can provide a sample enhanced in interactions on H.
For this sample, the purity for interactions on H is 27\%, which is an improvement on the 11\% seen with no \dt cut.

This study would benefit from a more sophisticated simulation that includes backgrounds and a better model for detector effects.
That is particularly true for the interactions on H since the limiting effects of FSI and nuclear recoil are not present.  For the larger targets, nuclear effects in the interaction model will inject uncertainty that needs to be estimated carefully.

\section{Conclusions}

The technique presented in Sec.~\ref{section:technique} to reconstruct neutrons on an event-by-event basis in neutrino interactions makes use of the 3D direction of the tagged neutron.
The study summarized here shows that the use of additional information provides superior performance in reconstructing the neutrino energy relative to purely calorimetric or missing transverse momentum balance methods.
The neutron kinetic energy resolution is more symmetric and peaks closer to the true value than what is seen assigning the missing transverse momentum to be the neutron momentum.
The details of the study are clearly model dependent through the use of the neutrino generator and assumptions about detector effects.  Still, studies done varying the detection thresholds, choosing to select subleading neutron candidates in some fraction of the events, and using a different neutrino interaction generator showed the results to be robust, implying that the basic conclusions are sound. 

The study shows that the region of small \dt can be selected to enhance the purity of interactions on H in a plastic scintillator target.
Also a cut on \dt is potentially useful to enhance the signal to noise for neutron candidates. 

The technique is potentially useful for any detector that is able to associate a neutron with a given neutrino interaction and determine the neutron direction. If a detector can tag the 3D position of a neutron energy deposition created as it propagates this condition is satisfied.
This would include LArTPC detectors which have a demonstrated capability to detect proton knockout interactions by neutrons and the SuperFGD detector in the T2K upgrade.
At higher \Tn, where the neutron produces a larger energy deposition, the 3D position of the deposition can be determined in strip-based scintillator detectors as well, such as MINERvA. 

For the T2K SuperFGD, the Time-of-Flight between the neutrino interaction and the neutron energy deposition can be used to determine \Tn.  In such a detector, the usefulness of the technique presented here is that it is much less dependent on the lever arm and timing resolution.   This technique may prove more or less useful than the ToF technique depending on the event morphology and the expected timing resolution of the given neutron candidate.

For LArTPC's this technique offers a way to include a fraction of the produced neutrons in the event-by-event reconstruction that has, to date, been lacking.  For antineutrino-induced events in particular, this technique may provide a sample with an improved neutrino energy resolution.


\section{Acknowledgements}
This work has been supported by the U.S. Department of Energy.
Additional support came from EURO-LABS, which received funding from the European Union’s Horizon Europe Research and Innovation programme under grant agreement No 101057511.

\bibliography{biblio}

@article{genie0,
    author = "Andreopoulos, C. and others",
    title = "{The {GENIE} Neutrino Monte Carlo Generator}",
    eprint = "0905.2517",
    archivePrefix = "arXiv",
    primaryClass = "hep-ph",
    reportNumber = "FERMILAB-PUB-09-418-CD",
    doi = "10.1016/j.nima.2009.12.009",
    journal = "Nucl. Instrum. Meth. A",
    volume = "614",
    pages = "87--104",
    year = "2010"
}

@article{nuis:2016jfr,
  author = "Stowell, P. and others",
  title = "{NUISANCE: a neutrino cross-section generator tuning and comparison framework}",
  eprint = "1612.07393",
  archivePrefix = "arXiv",
  primaryClass = "hep-ex",
  doi = "10.1088/1748-0221/12/01/P01016",
  journal = "JINST",
  volume = "12",
  number = "01",
  pages = "P01016",
  year = "2017"
}

@article{SoLid:2017ema,
   title="{A novel segmented-scintillator antineutrino detector}",
   collaboration={SoLid Collaboration},
   volume={12},
   ISSN={1748-0221},
   url={http://dx.doi.org/10.1088/1748-0221/12/04/P04024},
   DOI={10.1088/1748-0221/12/04/p04024},
   number={04},
   journal={Journal of Instrumentation},
   publisher={IOP Publishing},
   author={Abreu, Y. and others},
   year={2017},
   month=apr,
   pages={P04024–P04024} }

@article{Super-Kamiokande:2021the,
title = "{First gadolinium loading to Super-Kamiokande}",
journal = {Nuclear Instruments and Methods in Physics Research Section A: Accelerators, Spectrometers, Detectors and Associated Equipment},
collaboration = {Super-Kamiokande Collaboration},
volume = {1027},
pages = {166248},
year = {2022},
issn = {0168-9002},
doi = {https://doi.org/10.1016/j.nima.2021.166248},
url = {https://www.sciencedirect.com/science/article/pii/S0168900221010883},
author = {K. Abe and others}
}

@misc{ANNIE:2017nng,
      title="{Accelerator Neutrino Neutron Interaction Experiment ({ANNIE}): Preliminary Results and Physics Phase Proposal}", 
      collaboration={ANNIE Collaboration},
      author={A. R. Back and others},
      year={2017},
      eprint={1707.08222},
      archivePrefix={arXiv},
      primaryClass={physics.ins-det},
      url={https://arxiv.org/abs/1707.08222}, 
}

@article{Reines:1960pr,
  title = "{Detection of the Free Antineutrino}",
  author = {Reines, F. and others},
  journal = {Phys. Rev.},
  volume = {117},
  issue = {1},
  pages = {159--173},
  numpages = {0},
  year = {1960},
  month = {Jan},
  publisher = {American Physical Society},
  doi = {10.1103/PhysRev.117.159},
  url = {https://link.aps.org/doi/10.1103/PhysRev.117.159}
}

@article{MINERvA:2019wqe,
  title = "{Neutron measurements from antineutrino hydrocarbon reactions}",
  author = {Elkins, M. and others},
  collaboration = {MINERvA Collaboration},
  journal = {Phys. Rev. D},
  volume = {100},
  issue = {5},
  pages = {052002},
  numpages = {20},
  year = {2019},
  month = {Sep},
  publisher = {American Physical Society},
  doi = {10.1103/PhysRevD.100.052002},
  url = {https://link.aps.org/doi/10.1103/PhysRevD.100.052002}
}

@article{MINERvA:2023ikp,
  title = "{Measurement of the multineutron ${\overline{\ensuremath{\nu}}}_{\ensuremath{\mu}}$ charged current differential cross section at low available energy on hydrocarbon}",
  author = {Olivier, A. and others},
  collaboration = {$\mathrm{MINER}\ensuremath{\nu}\mathrm{A}$ Collaboration},
  journal = {Phys. Rev. D},
  volume = {108},
  issue = {11},
  pages = {112010},
  numpages = {12},
  year = {2023},
  month = {Dec},
  publisher = {American Physical Society},
  doi = {10.1103/PhysRevD.108.112010},
  url = {https://link.aps.org/doi/10.1103/PhysRevD.108.112010}
}

@Article{MINERvA:2023avz,
author={Cai, T. and others},
title="{Measurement of the axial vector form factor from antineutrino-proton scattering}",
journal={Nature},
year={2023},
month={Feb},
day={01},
volume={614},
number={7946},
pages={48-53},
issn={1476-4687},
doi={10.1038/s41586-022-05478-3},
url={https://doi.org/10.1038/s41586-022-05478-3}
}

@phdthesis{ref:LastThesis,
author = {Last, D.},
title = "{Effect of Nuclear Size on Energetic Neutron Production in CCQE-Like Antineutrino Interactions in the MINERvA Detector}",
school = {University of Pennsylvania},
year = {2024},
url = {https://ezp.lib.rochester.edu/login?url=https://www.proquest.com/dissertations-theses/effect-nuclear-size-on-energetic-neutron/docview/3068748669/se-2?accountid=13567}
}

@misc{T2K:2019bbb,
      title="{{T2K ND280} Upgrade -- Technical Design Report}", 
      author={K. Abe and others},
      collaboration = {T2K Collaboration},
      year={2020},
      eprint={1901.03750},
      archivePrefix={arXiv},
      primaryClass={physics.ins-det},
      url={https://arxiv.org/abs/1901.03750}, 
}

@article{Munteanu:2019llq,
  title = "{New method for an improved antineutrino energy reconstruction with charged-current interactions in next-generation detectors}",
  author = {Munteanu, L. and others},
  journal = {Phys. Rev. D},
  volume = {101},
  issue = {9},
  pages = {092003},
  numpages = {12},
  year = {2020},
  month = {May},
  publisher = {American Physical Society},
  doi = {10.1103/PhysRevD.101.092003},
  url = {https://link.aps.org/doi/10.1103/PhysRevD.101.092003}
}

@misc{Hyper-Kamiokande:2018ofw,
      title="{{Hyper-Kamiokande} Design Report}", 
      author={K. Abe and others},
      collaboration = {Hyper-Kamiokande Proto-Collaboration},
      year={2018},
      eprint={1805.04163},
      archivePrefix={arXiv},
      primaryClass={physics.ins-det},
      url={https://arxiv.org/abs/1805.04163}, 
}

@article{MicroBooNE:2024hun,
	author = {Abratenko, P. and others},
    collaboration = {MicroBooNE Collaboration},
	title = "{Demonstration of neutron identification in neutrino interactions in the MicroBooNE liquid argon time projection chamber}",
	DOI= "10.1140/epjc/s10052-024-13423-z",
	url= "https://doi.org/10.1140/epjc/s10052-024-13423-z",
	journal = {Eur. Phys. J. C},
	year = 2024,
	volume = 84,
	number = 10,
	pages = "1052",
}

@article{myatt,
  author = {Myatt, G.},
  title = "{The Use of Transverse Momentum Balance as a Means of Estimating the Energy of Neutrino Interactions, contribution to the Proceedings of the ECFA 300 GeV Working Group, Volume II, p. 117}",
  journal = {CERN/EFCA/72-4},
  year = {1972},
  month={September},
  note={},
}

@article{nd280flux,
  title = "{{T2K} neutrino flux prediction}",
  author = {Abe, K. and others},
  collaboration = {T2K Collaboration},
  journal = {Phys. Rev. D},
  volume = {87},
  issue = {1},
  pages = {012001},
  numpages = {34},
  year = {2013},
  month = {Jan},
  publisher = {American Physical Society},
  doi = {10.1103/PhysRevD.87.012001},
  url = {https://link.aps.org/doi/10.1103/PhysRevD.87.012001}
}

@misc{DUNE:2021cuw,
      title="{Experiment Simulation Configurations Approximating DUNE TDR}", 
      author={B. Abi and others},
      collaboration = {DUNE Collaboration},
      year={2021},
      eprint={2103.04797},
      archivePrefix={arXiv},
      primaryClass={hep-ex},
      url={https://arxiv.org/abs/2103.04797}, 
}


\end{document}